\documentclass[twoside,12pt]{article}
\usepackage{epsfig}

\def\Journal#1#2#3#4{{#1} {#2} (#4) #3 }

\def\NPA{{\em Nucl. Phys.} A}

\def\PRL{\em Phys. Rev. Lett.}

\def\PRC{{\em Phys. Rev.} C}

\def\INT{{\em Int. J. Mod. Phys.} E}

\newcommand{\be}{\begin{equation}}
\newcommand{\ee}{\end{equation}}
\newcommand{\bea}{\begin{eqnarray}}
\newcommand{\eea}{\end{eqnarray}}

\begin{document}

\title{ \vspace{1cm} Experimental approaches for determining in-medium properties of hadrons from photo-nuclear reactions}
\author{V.\  Metag,$^1$,\  M.\ Thiel$^{1,}$\footnote{present address: Institut f\"ur Kernphysik, Universi\"at Mainz}\ ,\  H.\ Bergh\"auser$^1$,\  S.\ Friedrich$^1$,\  B.\ Lemmer$^{1,}$\footnote{present address: II. Physikalisches Institut, Universit\"at G\"ottingen}\ ,\\
U.\ Mosel$^2$, J. \ Weil$^2$, \\
for the A2 collaboration\\
$^1$ II. Physikalisches Institut, Universit\"at Giessen, Germany\\
$^2$ Institut  f\"ur Theoretische Physik, Universit\"at Giessen, Germany}
\maketitle

\begin{abstract} 
Properties of hadrons and their modification within strongly interacting matter provide a link between experimental
observables and Quantum Chromodynamics (QCD) in the non-perturbative sector. The sensitivity of various observables to in-medium modifications of mesons is discussed. The transparency ratio, comparing the meson yield per nucleon within a nucleus relative to that on a free nucleon, is related to the in-medium width of the meson. While the transparency ratio can be determined for any meson lifetime the meson line shape only contains information on in-medium properties if the meson is so short-lived that it decays in the medium after production in a nuclear reaction. Light vector mesons are thus particularly suited for these investigations. The momentum distribution of mesons produced in a photo-nuclear reaction as well as the excitation function also show some sensitivity to different in-medium modification scenarios. As an example, high statistics data taken at MAMI-C on the photoproduction of 
$\omega$ mesons are presented. 
 \end{abstract}

\section{Introduction}
In-medium changes of hadron properties are one of the key problems in Quantum Chromodynamics (QCD) in the strong coupling regime. The question is how experimentally well known hadron spectra change when these hadrons are embedded in a strongly interacting environment. Spectral modifications of hadrons, encoded in changes of their mass and decay width, are often discussed in the context of a restoration of the the broken chiral symmetry in nuclei. 
The links between hadron properties and QCD symmetries are, however,  much more involved than initially thought, as discussed in  \cite{Leupold:2009kz}. Thus, hadronic models, based on our current understanding of meson-baryon interactions, have been used by several theory groups to calculate the in-medium self-energies of hadrons and their spectral functions. 
\begin{figure}[tb]
\begin{center}
\begin{minipage}[t]{16 cm}
\epsfig{file=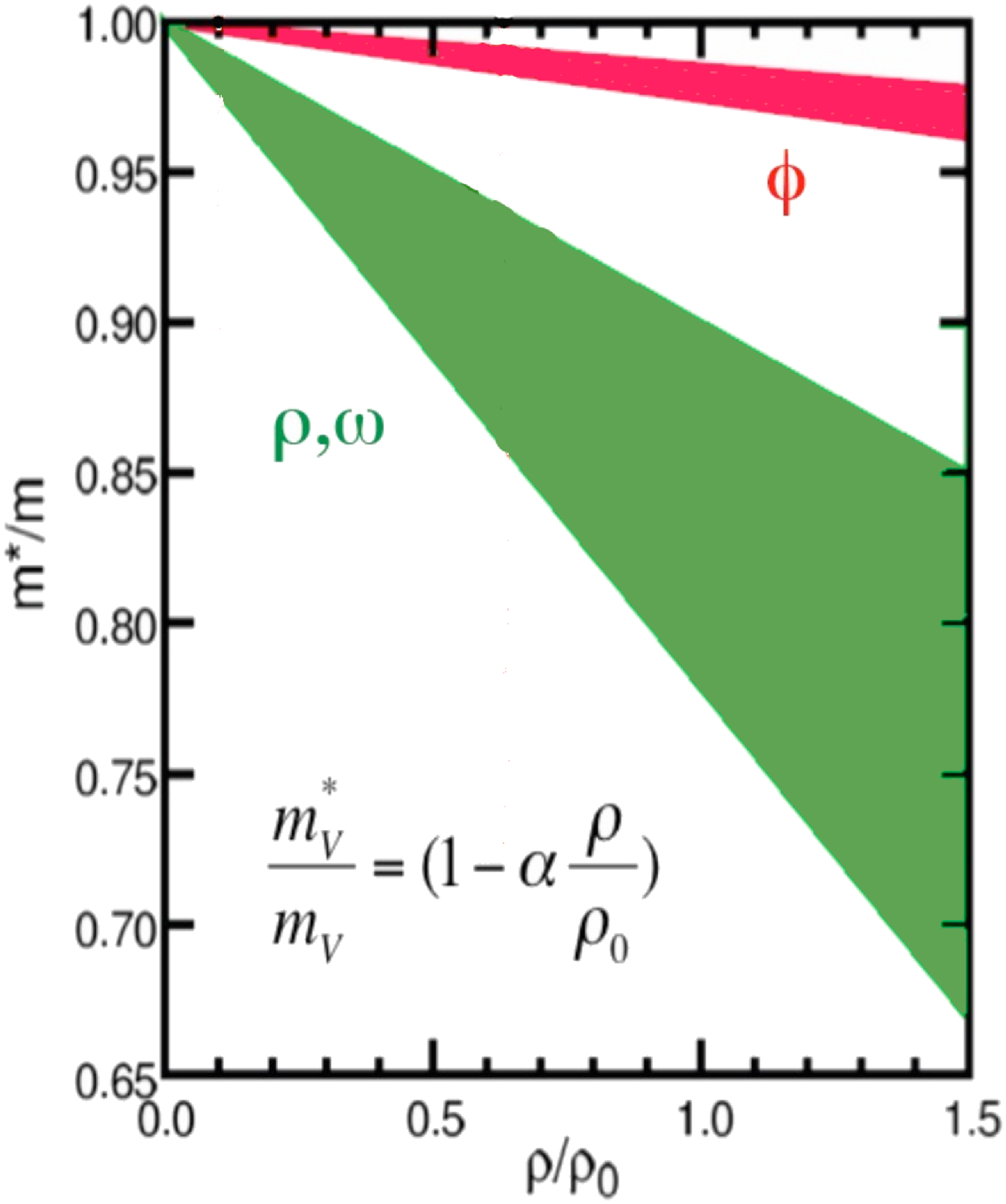, scale=0.18}
\epsfig{file=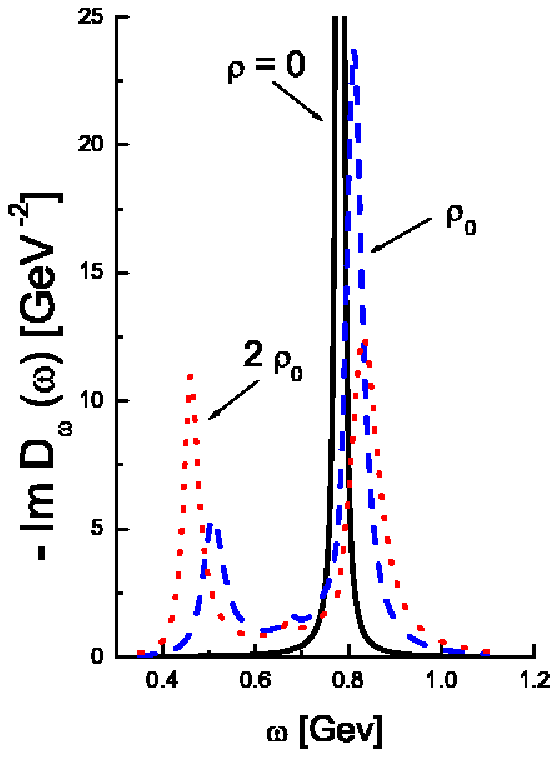,width=5.0cm,totalheight=5.0cm}
\epsfig{file=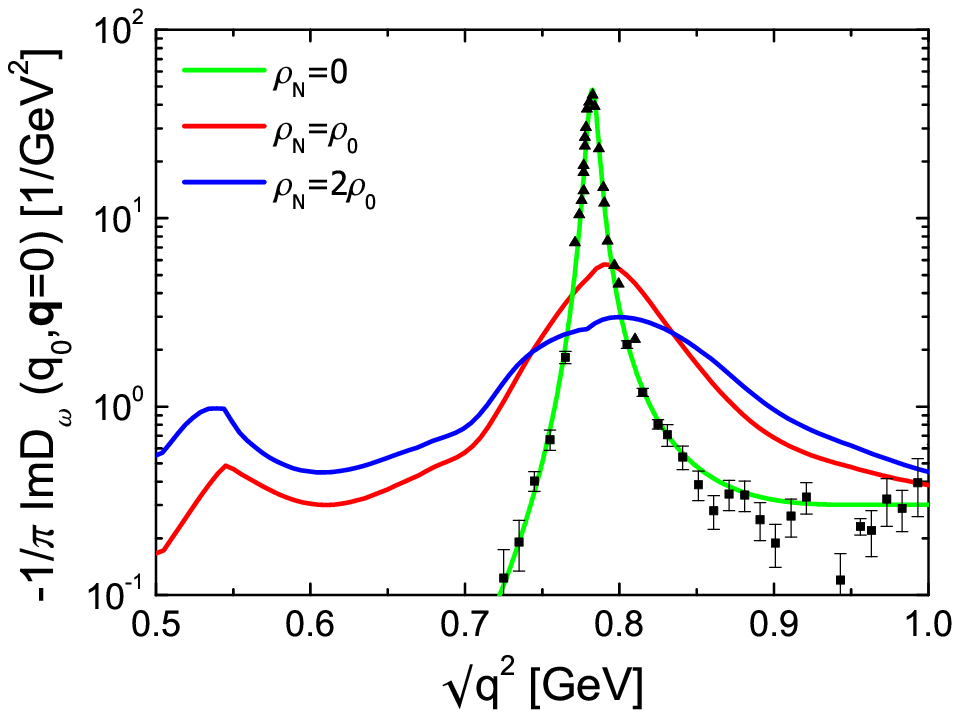,scale=0.65}
\end{minipage}
\begin{minipage}[t]{16.5 cm}
\caption{Predictions for in-medium modifications of vector mesons: (Left) Decrease of vector-meson masses with increasing nuclear density as predicted by Hatsuda and Lee \cite{Hatsuda:1991ez} using QCD sum rules. (Middle) Structures in the $\omega$ spectral function arising from the coupling of the $\omega$ meson to nucleon resonances \cite{Lutz}. (Right) In-medium broadening of the $\omega$ meson calculated in a coupled channel approach \cite{Muehlich}.\label{fig:theo}}
\end{minipage}
\end{center}
\end{figure}
As an example, Fig.~\ref{fig:theo} illustrates different in-medium modification scenarios which have been predicted. Applying a QCD sum-rule approach Hatsuda and Lee \cite{Hatsuda:1991ez} predict a lowering of vector meson masses by 10-20$\%$ at normal nuclear-matter density for $\rho$ and $\omega$ mesons while for the $\phi$ meson only a small mass drop is expected due to the only weak interaction of $\phi$ mesons with nuclei. Lutz et al. \cite{Lutz} calculate a spectral function of the $\omega$ meson exhibiting a slight upward shift in mass and a structure near 500 MeV/c$^2$ arising from the coupling of the $\omega$ meson to nucleon resonances. The calculation by M\"uhlich et al. \cite{Muehlich} exhibits a strong broadening of the $\omega$ meson at normal and twice normal nuclear-matter density. These calculations refer to mesons at rest in the nucleus. An experimental test of these theoretical predictions thus requires measurements sensitive to mass shifts, structures and/or  broadening of hadronic spectral functions and detectors with acceptance for low momentum mesons. With this in mind, a series of photo-nuclear experiments has been performed which are described in the following sections.

\section{Experimental Approaches and Results}\label{expapp}
In recent years, several approaches have been developed to extract in-medium properties of mesons from experimental data:

\begin{figure}[t]
\begin{center}
\begin{minipage}[t]{16 cm}
\epsfig{file=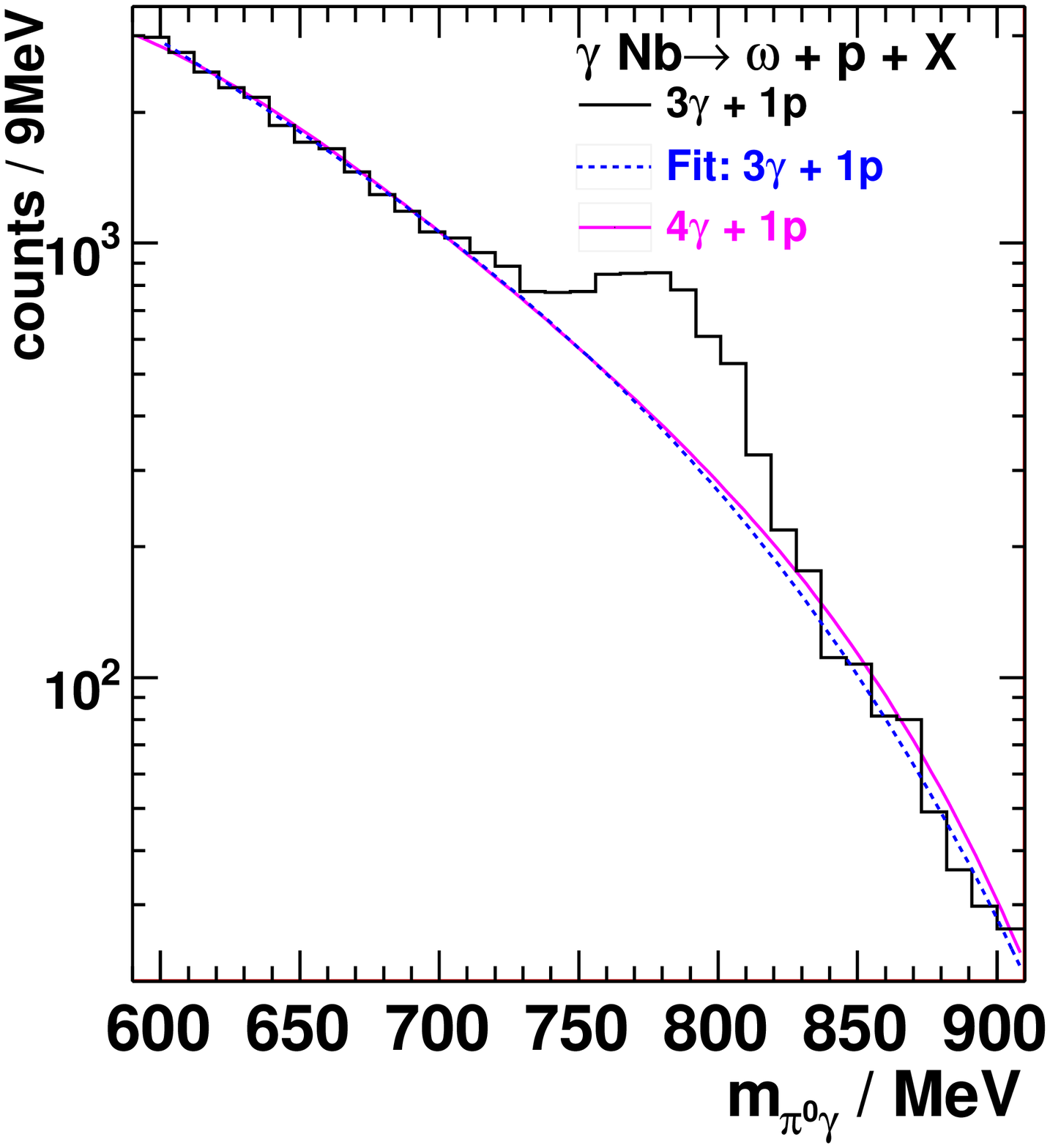,scale=0.30}
\hspace*{2cm}
\epsfig{file=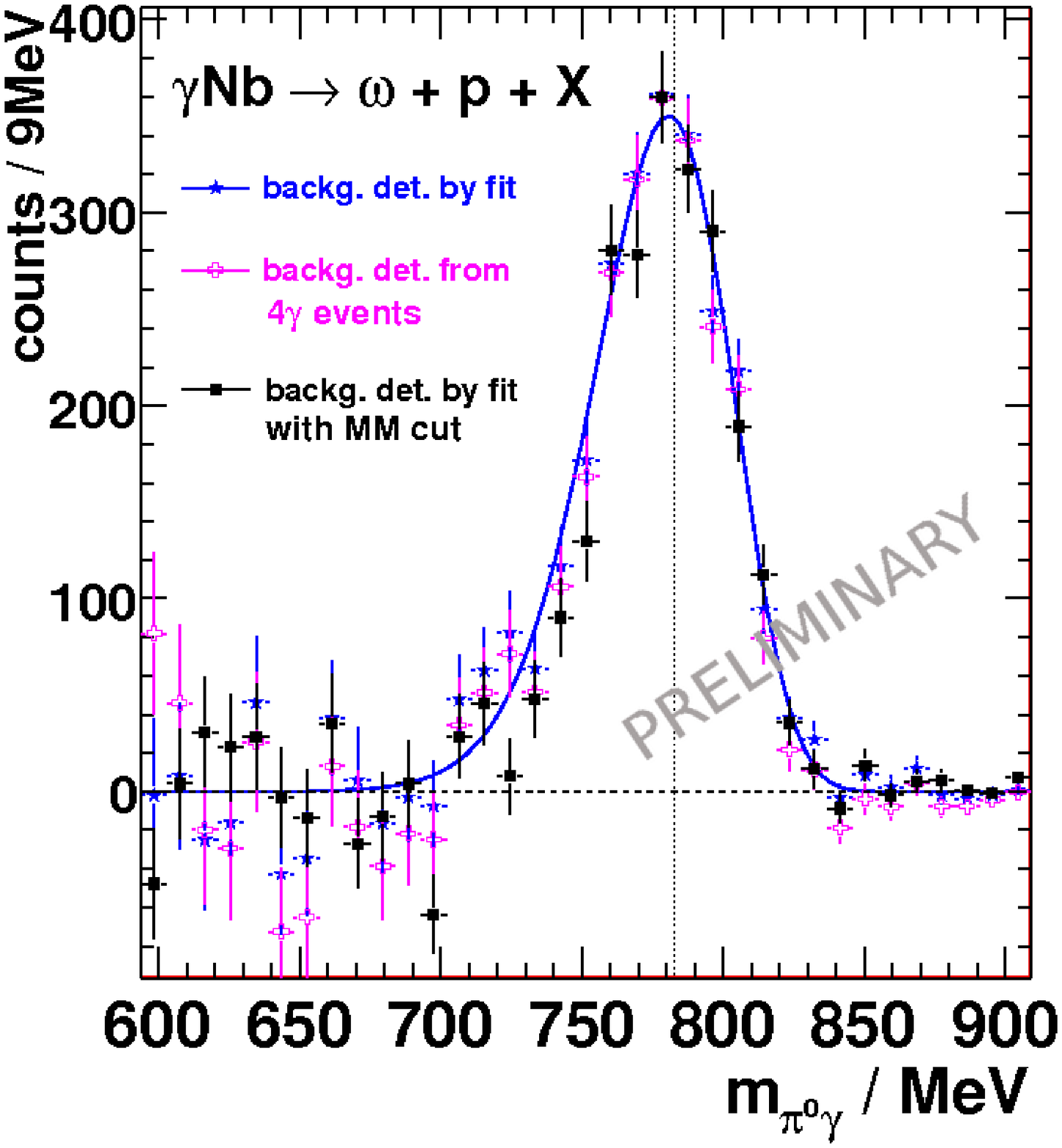,scale=0.32}
\end{minipage}
\begin{minipage}[t]{16.5 cm}
\caption{Left: Invariant mass spectrum of $\pi^0 \gamma$ pairs for photons of 900-1300 MeV incident on a Nb target for events with 3 photons and 1 proton registered in the Crystal Ball/TAPS detector system. The dotted (blue) curve is a fit to the background. The solid (red) curve represents the background determined from events with 4 photons and 1 proton, following the procedure described in \cite{Nanova:2010sy}. Right:  
$\pi^0 \gamma$ invariant mass spectrum after subtracting the different background shapes. The black data points represent the $\omega$ line shape after applying an additional missing mass cut. \label{fig:pig}}
\end{minipage}
\end{center}
\end{figure}

\begin{figure}[t]
\begin{center}
\begin{minipage}[t]{16 cm}
\epsfig{file=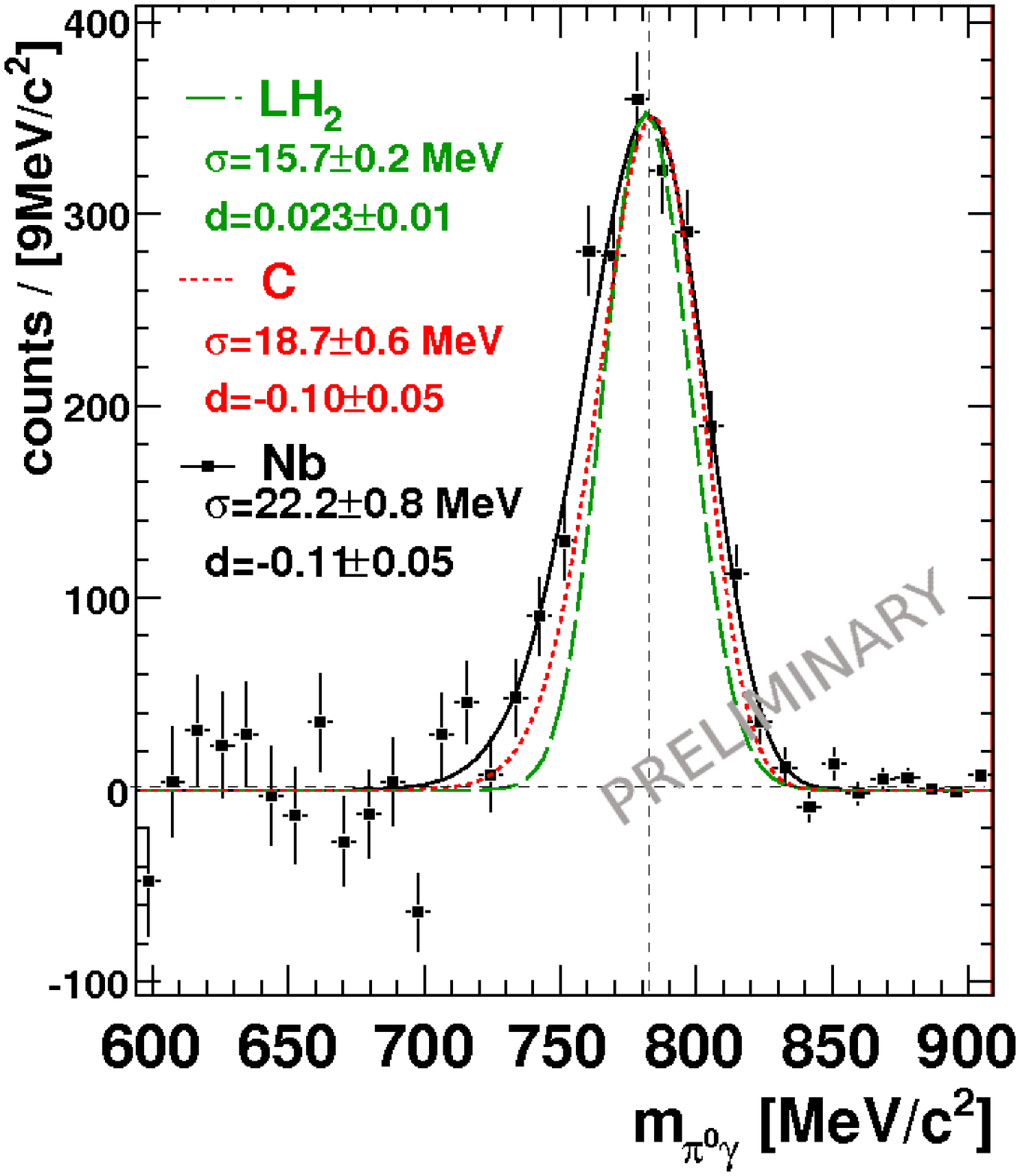, scale=0.3}
\hspace*{2cm} 
\epsfig{file=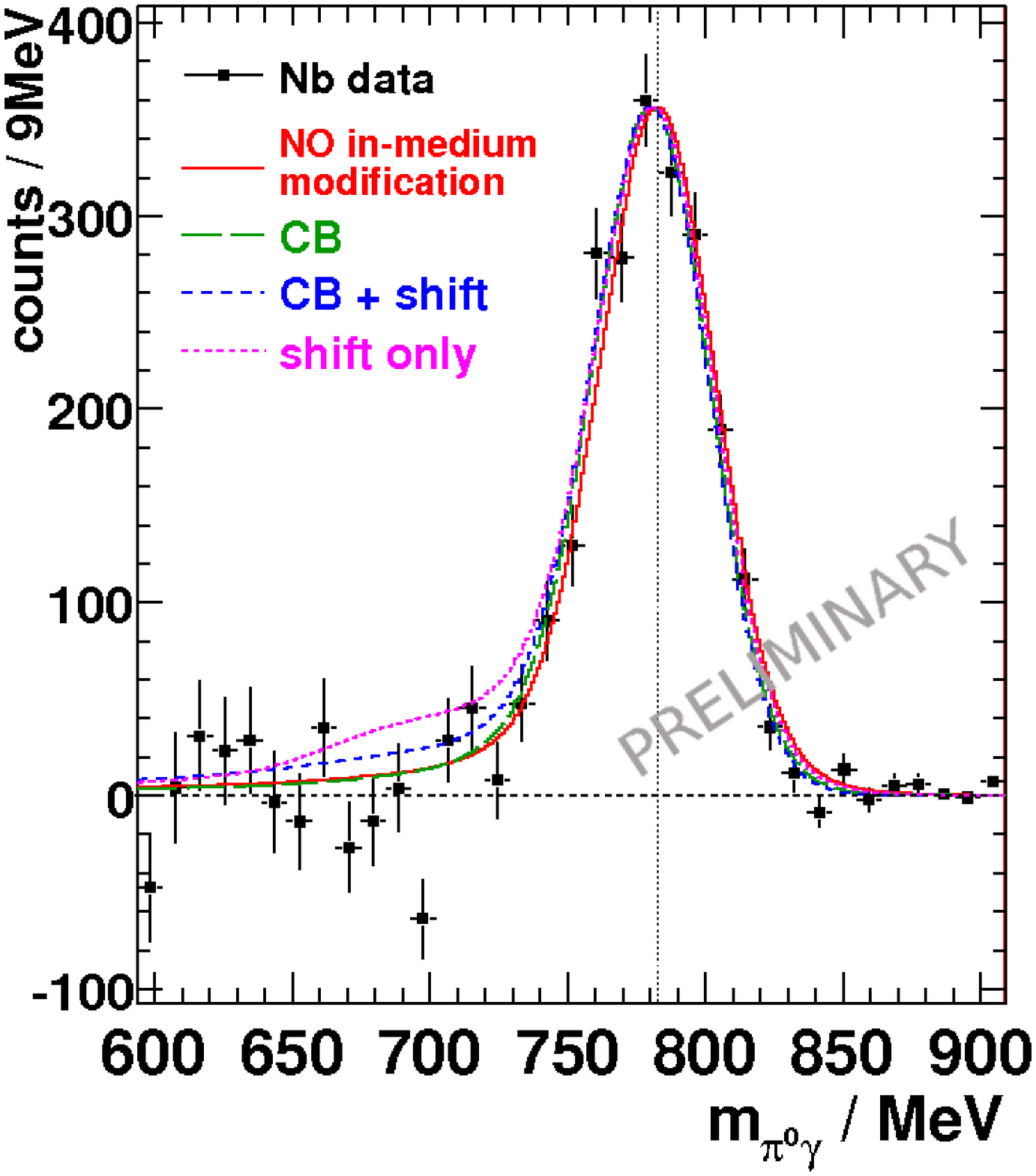,scale=0.3}
\end{minipage}
\begin{minipage}[t]{16.5 cm}
\caption{Left: The $\pi^0 \gamma$ invariant mass spectrum of Fig.~\ref{fig:pig} right with missing mass cut in comparison to the corresponding spectra for a C and a LH2 target. The distortion of the signal width due to the finite target length, determined in simulations, has been corrected for. Right: Comparison of the observed  $\omega$  line shape with GiBUU calculations \cite{GiBUU,Nanova_EPJA} for different in-medium modification scenarios, assuming no in-medium modifications (red solid curve), only collisional broadening (green dashed curve), collisional broadening and mass shift by -16$\%$ at normal nuclear matter density (short dashed, blue curve), and mass shift without broadening (dotted, magenta curve).\label{fig:omega_compare}}
\end{minipage}
\end{center}
\end{figure}

\subsection{Analysis of meson line shape}

The in-medium mass $\mu$ of the meson can 
be deduced from the 4-momentum vectors $p_1,p_2$ of the decay products
according to
\begin{equation}
\mu(\vec{p},\rho) = \sqrt{(p_1 + p_2)^2} \,. \label{VM:mass}
\end{equation}
In general, the mass distribution $\mu(\vec{p},\rho)$ depends on the
3-momentum $\vec{p}$ of the vector meson and on the density $\rho$ of the nuclear medium at the decay point. An in-medium mass shift of a meson would become evident by comparing the mass calculated from Eq.~1 in the limit of low meson momenta with the vacuum mass of this meson listed in \cite{PDG}. 

The light vector mesons $\rho,\omega,$ and $\phi$ are particularly suited for the mass distribution measurements
since their lifetimes of 1.3 fm/c, 23 fm/c and 46 fm/c, respectively, are so short that they decay within the nuclear medium with some probability after production in a nuclear reaction. Nevertheless, severe momentum cuts have to be applied for the longer lived $\omega$ and $\phi$ mesons to achieve decay lengths comparable to nuclear dimensions. 

Calculating the in-medium mass of the meson from Eq.~1 implies that the 4-momentum vectors of the decay products are not distorted by final state interactions with the surrounding nuclear medium. Dileptons are thus the optimum decay channel as any final-state distortion of the 4-momenta of the
decay products entering Eq.\ref{VM:mass} is avoided. Unfortunately, the branching ratios into the dilepton channels are only of the order of
$10^{-5}$-$10^{-4}$, making these measurements very difficult and sensitive to background subtraction. In this work, the $\omega \rightarrow \pi^0 \gamma$ decay with a branching ratio of 8.9$\%$ is utilized. As shown in \cite{Messchendorp:2001pa,Kaskulov}, the final-state interaction of the $\pi^0$ meson can be sufficiently suppressed by applying a cut on the kinetic energy of the $\pi^0$ mesons $T_{\pi} \ge 150 $MeV.

As pointed out in \cite{Eichstaedt}, any measurement of a mass distribution in a photo-nuclear reaction does, however, not yield the hadronic spectral function directly but rather a convolution with the branching ratio $\Gamma_{V \to  p_1 + p_2}/\Gamma_{\rm tot}$ into the specific final channel. Any increase in the total width $\Gamma_{\rm tot}$ in the hadronic environment leads to a reduction of this branching ratio. Thus, the strong broadening of the $\omega$ meson due to inelastic reactions within the nuclear medium, observed in \cite{Kotulla:2008xy} and discussed below,  suppresses the decay modes available to the free 
$\omega$ meson. This lowers the fraction of in-medium  $\omega \rightarrow \pi^0 \gamma$ decays and reduces the sensitivity of the line-shape analysis to in-medium modifications. Furthermore, ambiguities in the subtraction of the background in $\pi^0 \gamma$ invariant mass spectra leads to additional uncertainties in the determination of the $\omega$ line shape, as discussed in \cite{Kaskulov,Trnka,Nanova:2010sy,Nanova_EPJA}. In a series of experiments with tagged photon beams at MAMI-C, using the Crystal Ball/ TAPS detector system, an attempt has been made to reduce the statistical and systematic errors in the $\omega$ line shape determination. Experimental details are given in \cite{Micha_PhD}. The $\pi^0 \gamma$ invariant-mass spectrum obtained in $\omega$ photoproduction off Nb is shown in Fig.~ \ref{fig:pig}. The 
$\omega$ peak resides on a background mostly from 2$\pi^0$ and $\pi^0 \eta$ events where one of the 4 decay photons was not registered in the  Crystal Ball/TAPS detector system. The background has been determined by a.) a fit with a 5th order polynomial and b.) a method developed by Nanova et al. \cite{Nanova:2010sy}, where the background is derived from the measured 2$\pi^0$ and $\pi^0 \eta$ events. Fig.\ref{fig:pig} right shows the $\omega$ signal after subtraction of the different background shapes. Furthermore, a third line-shape distribution resulting from an additional missing mass cut is also shown. Fig.\ref{fig:pig} right thus illustrates the systematic uncertainties associated with the different background treatments.

\begin{figure}[tb]
\begin{center}
\begin{minipage}[t]{16 cm}
\hspace*{2.5cm}
\epsfig{file=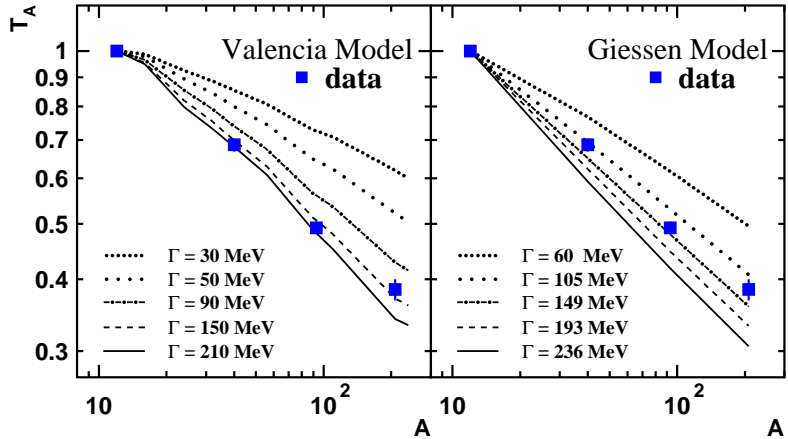,scale=0.55}
\end{minipage}
\begin{minipage}[t]{16.5 cm}
\caption{Transparency ratios for Ca, Nb, and Pb nuclei, experimentally determined according to Eq. \ref{T_A} and normalized to carbon. The data are compared (left) with Monte Carlo simulations \cite{Kaskulov:2006zc} and (right) BUU transport calculations \cite{Muehlich:2006ps} for different in-medium 
$\omega$ widths. \label{fig:T_A}}
\end{minipage}
\end{center}
\end{figure}

\begin{figure}[t]
\begin{center}
\begin{minipage}[t]{16 cm}
\hspace*{3cm}
\epsfig{file=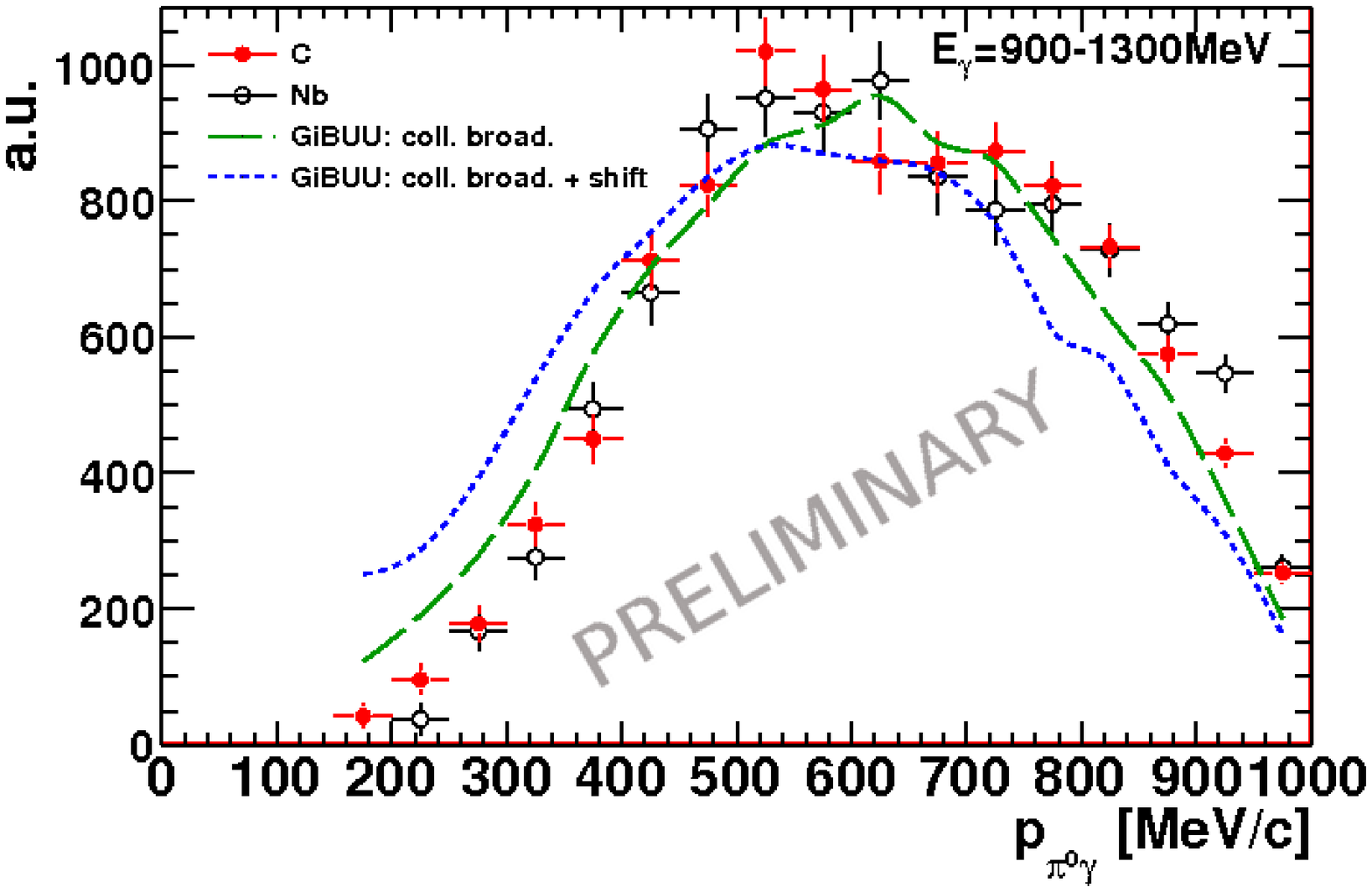,scale=0.5}
\end{minipage}
\begin{minipage}[t]{16.5 cm}
\caption{Momentum distribution of $\omega$ mesons for photons of 900 - 1300 MeV incident on C (open black circles) and Nb (closed red circles) targets. 
The preliminary data are compared to GiBUU transport model calculations, assuming collisional broadening of the $\omega$ meson with and without an in-medium mass shift of -16$\%$ at normal nuclear-matter density, respectively. \label{fig:momdist}}
\end{minipage}
\end{center}
\end{figure}
In Fig.~\ref{fig:omega_compare} the $\omega$ line shape obtained with an additional missing mass cut for the 1mm thick Nb target is compared to the 
corresponding $\omega$ signals for a C and a LH$_2$ target, respectively. For a quantitative comparison, the latter signals have been corrected for the loss in mass resolution due to the finite lengths of the targets of 15 and 50 mm, respectively. The signal on Nb is slightly broader than that for C and the reference signal obtained with the LH$_2$ target, consistent with the in-medium broadening of the $\omega$ meson found in the transparency-ratio measurement \cite{Kotulla:2008xy}, see below. 
\begin{figure}[tb]
\begin{center}
\begin{minipage}[t]{16 cm}
\epsfig{file=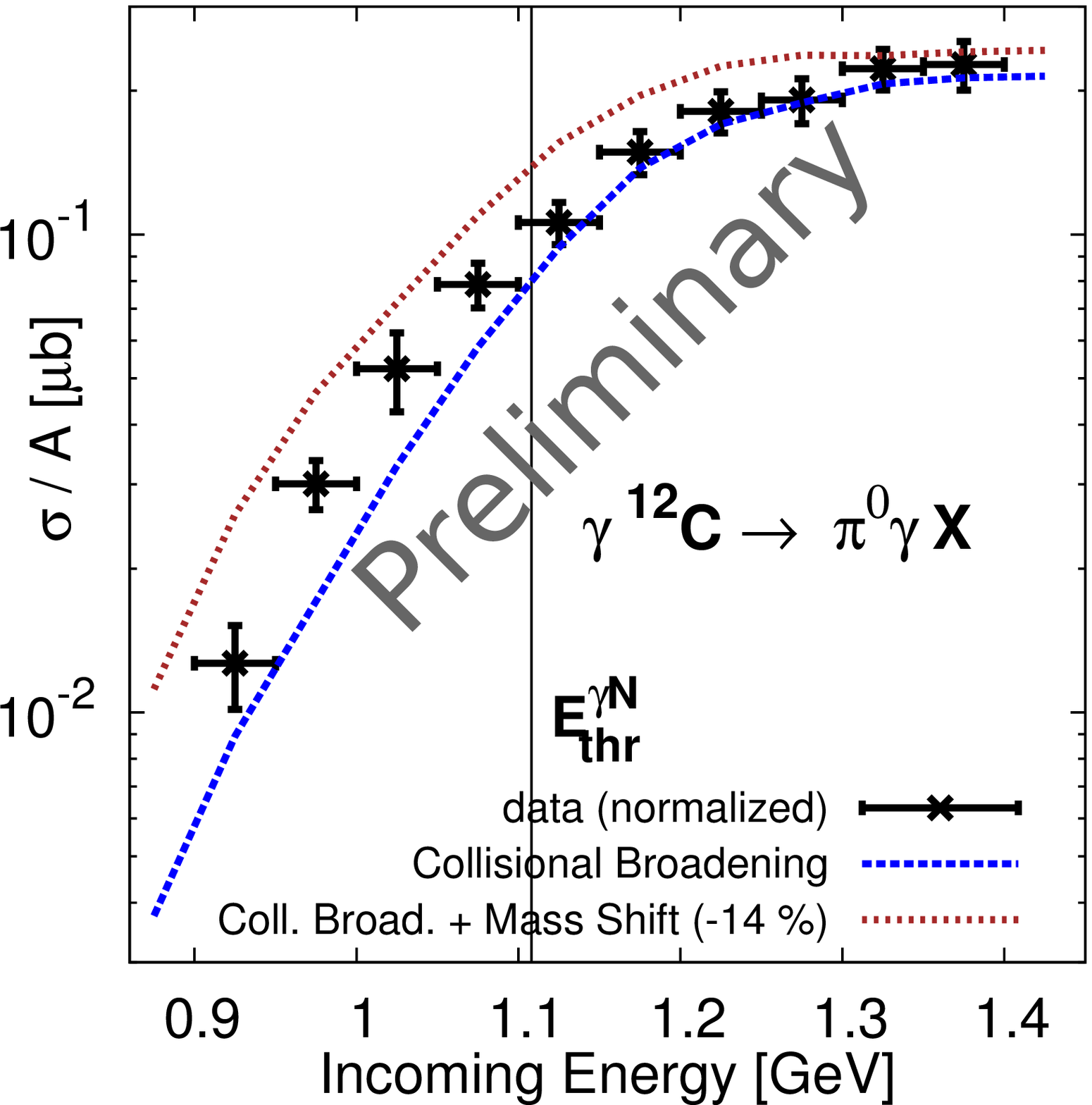,scale=0.35}
\hspace{2cm}
\epsfig{file=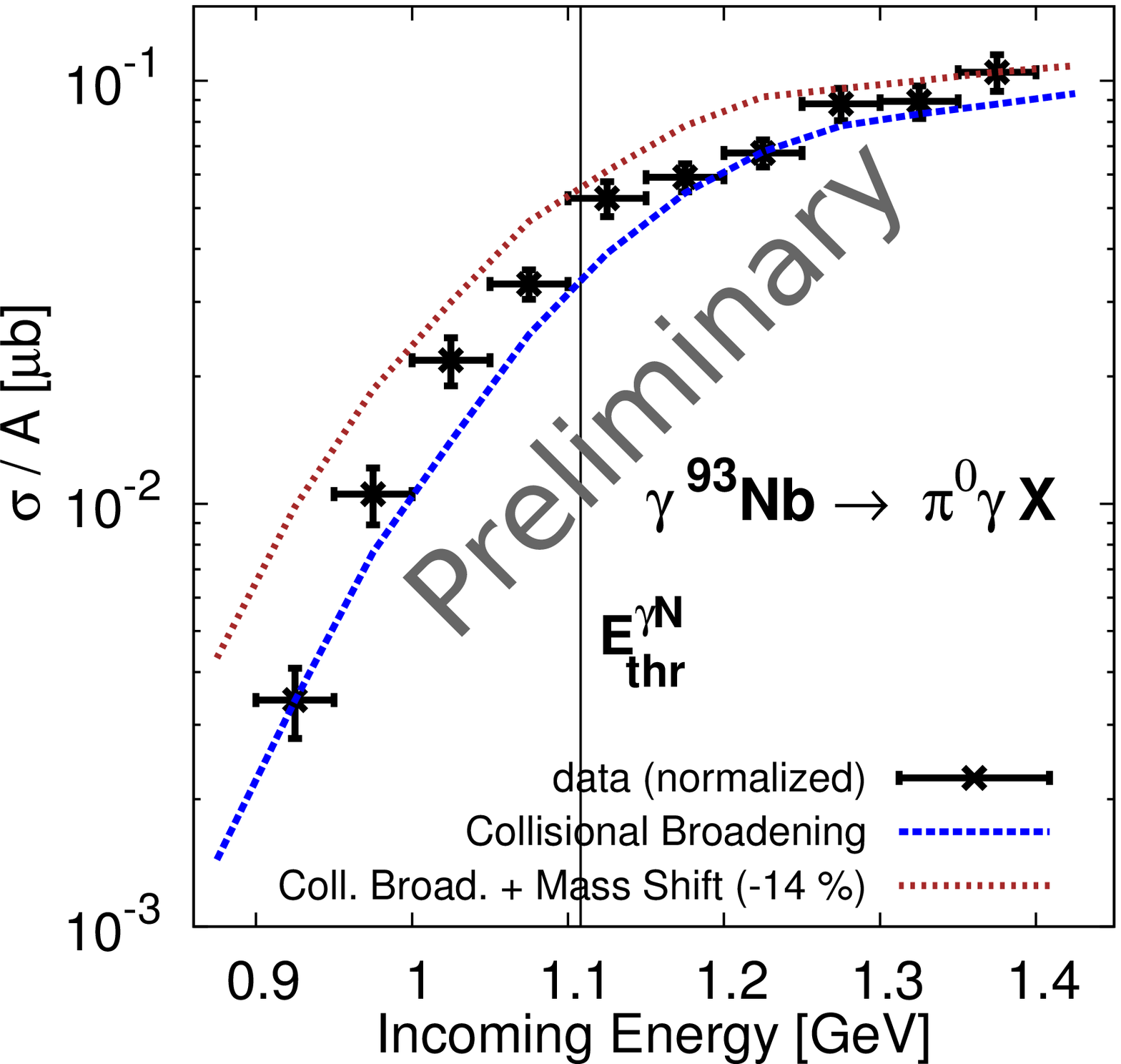,scale=0.35}
\end{minipage}
\begin{minipage}[t]{16.5 cm}
\caption{Excitation functions for photoproduction of $\omega$ mesons off C (left) and Nb (right), respectively. The vertical lines indicate the threshold energy for $\omega$ photoproduction on a free nucleon. The preliminary data are compared to GiBUU transport model calculations, assuming collisional broadening of the $\omega$ meson with and without an in-medium mass shift of -16$\%$ at normal nuclear-matter density, respectively. The preliminary data are normalized to the calculations at an incident photon energy of 1.375 GeV. 
\label{fig:exfunct}}
\end{minipage}
\end{center}
\end{figure}

In Fig.~\ref{fig:omega_compare} right the $\omega$ signal on Nb is compared to GiBUU transport calculations \cite{GiBUU} for different in-medium scenarios \cite{Nanova_EPJA}. Although the statistics of the experiment has been considerably improved, compared to previous studies 
\cite{Trnka,Nanova:2010sy,Nanova_EPJA},  it is still not sufficient to distinguish different scenarios which differ only very little in the predicted line shape. Only the case with mass shift and no broadening appears not be consistent with the data.

\subsection{In-medium $\omega$ width from attenuation measurements}

Access to the imaginary part of the in-medium self energy of hadrons is provided by attenuation measurements on nuclei with different mass number. In a 
nuclear medium, mesons can be removed by inelastic reactions with neighboring hadrons. As a consequence, the effective lifetimes of these mesons in the medium are shortened and their widths are correspondingly increased. The absorption of mesons can be extracted from the measurement of
the transparency ratio \protect\cite{Kaskulov:2006zc,Muehlich:2006ps}
\begin{equation}
T=\frac{\sigma_{\gamma A \rightarrow \omega X}}{A \cdot \sigma_{\gamma N \rightarrow \omega X}}.\label{T_A}
\end{equation}
The cross section for meson production per nucleon within a  
nucleus is compared with the meson production cross section on a free nucleon. The
nucleus serves as a target and at the same time also as an absorber. If there 
were no meson absorption in nuclei this ratio would be one.  
Since the meson photoproduction cross sections on the neutron are
not known in many cases, the transparency ratio is frequently normalized to the transparency ratio measured on a light nucleus like carbon. This normalization also helps to suppress the distortion of the transparency ratio by two-step production processes.

The transparency ratio for $\omega$ mesons has been measured in photoproduction experiments for several nuclei in a series of measurements with the CBELSA/TAPS detector system in Bonn. The result is shown in Fig.~\ref{fig:T_A} \cite{Kotulla:2008xy}. The $\omega$ yield per nucleon within a nucleus decreases with increasing nuclear mass number and drops to about 40$\%$ for Pb. A comparison of the data 
with calculations of the Valencia \protect\cite{Kaskulov:2006zc} 
and Giessen \cite{Muehlich:2006ps} theory groups (Fig.~\ref{fig:T_A})
indicates an in-medium $\omega$ width in the nuclear reference frame of about
130-150 MeV at normal nuclear-matter density for an average $\omega$
momentum of 1100 MeV/c. This implies an in-medium broadening of the $\omega$
meson by a factor $\approx$ 16. An even larger attenuation of $\omega$ mesons has recently been reported by the CLAS collaboration \cite{Wood:2010ei}.

\subsection{In-medium properties from measurement of momentum distributions}

Mesons produced in a nuclear reaction leave the nuclear medium with their free mass. In case of an in-medium mass drop, this mass difference has to be compensated at the expense of their kinetic energy.  As demonstrated in GiBUU  transport-model calculations this leads to a downward shift in the momentum distribution as compared to a scenario without mass shift. A mass shift can thus be indirectly inferred from a measurement of the momentum distribution of the meson. 

The momentum distribution for $\omega$ mesons for incident photon energies between 900 and 1300 MeV is shown in Fig.~\ref{fig:momdist}. The preliminary data points for C and Nb almost coincide. A comparison with GiBUU calculations clearly favors the collisional-broadening scenario without mass shift. Since the collisional broadening is experimentally established, only scenarios with collisional broadening are considered in this comparison.

\subsection{In-medium properties from measurement of the excitation function}

In case of an in-medium mass shift, the energy threshold for producing this meson is lowered, leading to an increase in phase space. As a consequence, the production cross section for a given incident beam energy will increase as compared to a scenario without mass shift (see \cite{Muehlich:2006ps}). This idea has already been exploited in attempts to derive in-medium properties of K- mesons in heavy-ion reactions \cite{Barth}.

Preliminary data on the cross section for $\omega$ photoproduction on C and Nb nuclei are shown  in Fig.~\ref{fig:exfunct}. The partial cross section for the $\omega \rightarrow \pi^0 \gamma $ decay is plotted as a function of the incident photon energy in comparison to GiBUU simulations for the 
collisional-broadening scenario with and without mass shift. Since the absolute experimental cross sections have not yet been determined, both data sets have been normalized to the respective calculated cross sections at an incident photon energy of 1.375 GeV. Again, the scenario without mass shift is favored.  A mass shift smaller than -16$\%$ can, however,  not be excluded from this comparison. 

\section{Conclusion}
Preliminary results on the line shape, momentum distribution and excitation function for $\omega$ photoproduction off C and Nb nuclei have been presented and discussed together with the previously established strong broadening of the $\omega$ meson at normal nuclear-matter density \cite{Kotulla:2008xy}.
Even with improved statistics an analysis of the $\omega$ line shape does not allow a discrimination between different in-medium modification scenarios.
Indirect approaches like analyses of the momentum distribution and the excitation function of $\omega$ mesons seem, however,  to favor the scenario of collisional broadening without mass shift. It is planned to extend these studies to other mesons, in particular to the $\eta^\prime$ meson \cite {Nanova_Erice}.
\\

We would like to thank M. Nanova for valuable comments. This work has been supported by DFG TR16 {\em "subnuclear structure of matter"}.

\end{document}